\documentclass[12pt,a4paper]{article}
\usepackage{graphicx}
\usepackage{color}
\usepackage{times}
\newcommand{\mdot}{\mbox{$\dot{M}$}}
\newcommand{\Rstar}{\mbox{$R_\ast$}}

\newcommand{\vinf}{\mbox{$v_\infty$}}
\newcommand{\xmm}{{\em XMM-Newton}}
\newcommand{\cxo}{{\em Chandra}}
\newcommand{\zpup}{\mbox{$\zeta$\,Pup}}
\newcommand{\zori}{\mbox{$\zeta$\,Ori}}

\newcommand{\Lbol}{\mbox{$L_{\rm bol}$}}
\newcommand{\Tx}{\mbox{$T_{\rm X}$}}
\newcommand{\Lx}{\mbox{$L_{\rm X}$}}

\newcommand{\myr}{\mbox{$M_\odot\,{\rm yr}^{-1}$}}

\def \etal   {\hbox{et~al.\/}}
\def\changed{}

\title{\bf X-rays, clumping and wind structures}
\author{Lidia Oskinova$^1$\thanks{lida@astro.physik.uni-potsdam.de}, 
Wolf-Rainer Hamann$^1$, 
Richard Ignace$^2$, and
Achim Feldmeier$^1$ \\
\vspace{1cm}\\
\normalsize $^1$ Universit\"at Potsdam, Germany\\ 
\normalsize $^2$ East Tennessee State University, TN, USA}
\date{\mbox{}}
\begin{document}
\maketitle
\pagestyle{empty}
%
%
\def\bull{\vrule height .9ex width .8ex depth -.1ex}
\makeatletter
\def\ps@plain{\let\@mkboth\gobbletwo
\def\@oddhead{}\def\@oddfoot{\hfil\tiny\bull\quad
``The multi-wavelength view of hot, massive stars''; 39$^{\rm th}$ Li\`ege Int.\ Astroph.\ Coll., 12-16 July 2010 \quad\bull}%
\def\@evenhead{}\let\@evenfoot\@oddfoot}
\makeatother
%
%
\def\beginrefer{\section*{References}%
\begin{quotation}\mbox{}\par}
\def\refer#1\par{{\setlength{\parindent}{-\leftmargin}\indent#1\par}}
\def\endrefer{\end{quotation}}
%
%
{\noindent\small{\bf Abstract:} 
X-ray emission is ubiquitous among massive stars. In the last decade,
X-ray observations revolutionized our perception of stellar winds but
opened a Pandora's box of urgent problems.  X-rays penetrating stellar
winds suffer mainly continuum absorption, which greatly simplifies the
radiative transfer treatment.  The small and large scale structures in
stellar winds must be accounted for to understand the X-ray emission
from massive stars.  The analysis of X-ray spectral lines can help to infer the
parameters of wind clumping, which is prerequisite for obtaining
empirically correct stellar mass-loss rates. The imprint of large
scale structures, such as CIRs and equatorial disks, on the X-ray
emission is predicted, and new observations are testing
theoretical expectations. The X-ray emission from magnetic stars proves to be
more diverse than anticipated from the direct application of the
magnetically-confined wind model. Many outstanding questions
about X-rays from massive stars will be answered when the models and
the observations advance.}

%
%
\section{Introduction}

Two aspects in studies of X-ray emission from massive stars attract
most attention: {\em i)} how X-rays are generated in massive stars,
and {\em ii)} how X-ray emission can be used in analyzing stellar
winds. In the basic concept, the wind has two components: a general
cool wind with temperature of $T_{\rm w}\sim 10$\,kK which contains
nearly all the wind mass, and a hot tenuous component with
$\Tx\sim$few\,MK where the X-rays originate. The X-ray photons suffer
continuum K-shell absorption in the cool wind and, in turn, can affect
the wind ionization via the Auger process.

In this review we concentrate on X-ray emission from single stars.
{\changed This is thermal emission from gases heated in the stellar wind
shocks or in magnetically confined wind regions. Cassinelli \& Olson
(1979) proposed X-radiation from a base coronal zone plus Auger
ionization in the surrounding cool wind to explain the superionzation
(e.g.\ N\,{\sc v}, O\,{\sc vi}) that was observed to be present in
Copernicus UV spectra of OB stars. From the analysis of {\em Einstein} 
spectra of OB-stars,  Cassinelli \& Swank (1983) concluded that the 
base corona idea was not correct since soft X-rays were observed.  The
Si\,{\sc xiii} and S\,{\sc xv}  line emission was detected  in the SSS
spectrum of the O-star $\zeta$ Ori.  These ions correspond to  high
temperature and are located at a energy where the wind would be thin to
X-rays. This led to a conclusion about two sources of X-ray emission,
X-rays that arise from fragmented shocks in the wind and X-rays from
very hot, probably magnetically confined loops, near the base of the
wind. Furthermore since X-ray variability was already known to be less
than about 1\%, Cassinelli \& Swank (1983) suggested that there had
to be thousands of shock fragments in the wind. }Radiation hydrodynamic
simulations of the nonlinear evolution of instabilities in stellar winds
were performed by Owocki, Castor, \& Rybicki (1988). They demonstrated
that the X-ray can originate from plasma heated by strong reverse
shocks, which arise when a high-speed, rarefied flow impacts on slower
material that has been compressed into dense shells. Feldmeier, Puls, \&
Pauldrach (1997) assumed a turbulent seed perturbation at the base of
the stellar wind and found that the shocks arising when the shells
collide are capable of explaining the observed X-ray flux. These 1D
hydrodynamical models predict plasma with temperatures 1--10\,MK which
is permeated with the cool wind.  X-rays suffer absorption as they
propagate outwards through the ensemble of dense, radially compressed
shells.


Waldron (1984) calculated the opacity of O-star winds for the X-ray 
radiation. The absorption of X-rays in Wolf-Rayet (WR) star winds was
investigated by Baum \etal\ (1992). They employed detailed non-LTE
stellar atmosphere models and showed that since the WR wind opacity is
very high, the observed X-rays must emerge from the far outer wind
region.  Hillier \etal\ (1993) computed the wind opacity of the O5Ia
star $\zeta$\,Pup. They found that the high opacity of the stellar wind
would completely block the soft X-rays ($<0.5$\,keV) unless some
significant fraction of hot plasma is located far out in the wind, at
distances exceeding 100\,\Rstar.

The shape of X-ray emission line profiles was predicted by MacFarlane
\etal\ (1991). They considered the effect of wind absorption on the emission
from an expanding shell of hot gas. When the cool wind absorption is
small, the line is broad and has a box-like shape. For stronger wind
absorption, the line becomes more skewed  (see Fig.\,7 in MacFarlane
\etal\ 1991). The line shape is largely determined  by a 
parameter $\tau_0$:   
\begin{equation} 
\tau_0=\kappa_\lambda R_\ast = \rho_{\rm
w}\chi_{\lambda} R_\ast 
\label{eq:t0}  
\end{equation}  
where $R_\ast$\,[cm] is the stellar radius, and the atomic opacity
$\kappa_\lambda$ is the product of the mass absorption coefficient
$\chi_{\lambda}$ [cm$^2$\,g$^{-1}$] and the density of the cool wind
($\rho_{\rm w}$) as defined from the continuity equation {\changed{
$\dot{M}=4\pi\rho_{\rm w} v(r)r^2R_\ast^2$, where $r$ is the radial
distance in units of $R_\ast$, and $v(r)$ is the velocity law, that can
be prescribed by the formula
$v(r)=v_\infty(1-1/r)^\beta$.}} MacFarlane \etal\ notice that when
$\tau_0$ increases, the red-shifted part of the line
($\Delta\lambda>0$) becomes significantly more attenuated than the
blue-shifted part.  They suggested that evaluating the line shape
can be used to determine $\tau_0$. {\changed{
The K-shell opacity varies with wavelength with the power between 2 and 3
(Hillier \etal\ 1993), therefore in the X-ray band $\tau_0$ should 
change by orders of magnitude. Consequently, the X-ray emission line 
shape at shorter and longer wavelengths should be different.}}
Waldron \& Cassinelli (2001) expanded the single-shock model of
MacFarlane \etal\ and considered emission from spherically symmetric
shocks equally distributed between 0.4\vinf\ and 0.97\vinf\, with
temperatures ranging from 2 to 10\,MK. In similar spirit, Ignace
(2001) provided a formalism that accounts for the emission from a flow
that is embedded with zones of X-ray emitting gas. Owocki \& Cohen
(2001) calculated model X-ray line profiles for various combinations of 
the parameters $\beta$, $\tau_0$, and onset radii for X-ray emission.

\section{The high-resolution X-ray spectra of O-type supergiants}

Waldron \& Cassinelli (2001) obtained the first high-resolution X-ray
spectrum of an O star. Their analysis of this \cxo\ spectrum of the O9.7Ib
star \zori\ revealed that the hot plasma is located relatively close
to the stellar core and that the line profiles appear to be symmetric,
and not skewed.

\begin{figure}[h]
\centering          
\includegraphics[width=6cm, angle=-90]{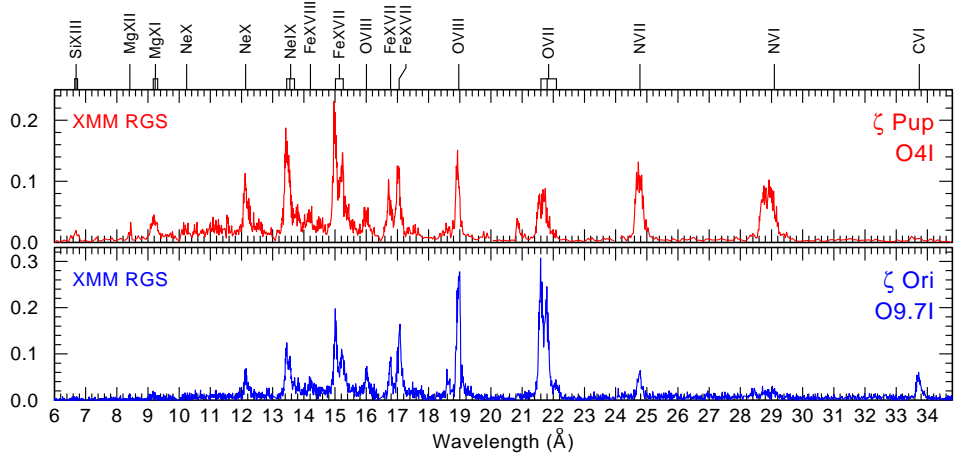}
\caption{RGS spectrum of the O4Ief star
$\zeta$\,Pup (top panel) and of the O9.7Ib star $\zeta$\,Ori (bottom panel),
not corrected for interstellar absorption. The 
$\zeta$\,Pup spectrum results from the combination of separate exposures
accumulating about 530 ks of useful exposure time. In the case of $\zeta$\,Pup, the
strength of the nitrogen lines compared to the oxygen and carbon lines
clearly indicates an overabundance of nitrogen. \label{fig:zxmm}}
\end{figure}

Subsequent analyses of X-ray spectra of single O-type stars are broadly
consistent with these first results. The \xmm\ RGS spectra of two O-type
stars are shown in Fig.\,\ref{fig:zxmm}. The general properties of X-ray
emission from massive stars are summarized in Waldron \& Cassinelli
(2007) and G\"udel \& Naz{\'e} (2009) (see also  Naz{\'e} 2011, these
proc.). The X-ray spectra of O-stars are well described by a thermal
plasma with temperatures spanning between $\approx 2-10$\,MK. The ratio
between the fluxes in forbidden and intercombination lines of He-like
ions indicates that the line formation region lies between $\approx
1.2-20\,\Rstar$ (e.g.\ Leutenegger \etal\ 2006). The line widths are
proportional to the terminal wind speed as obtained from UV line
diagnostics. The values of $\tau_0$ (see Eq.\,\ref{eq:t0}) are small and
the emission line profiles are similar across the X-ray spectrum.

\section{How to reconcile theory and observations}

The high-resolution X-ray spectra present two key problems. First, how
to explain the origin of X-rays at a distance of a few tens of stellar
radii from the photosphere? Second, how to explain the shape of the X-ray
emission lines and their similarity across the spectrum? 

The presence of magnetic fields may help to heat the plasma very close
to the stellar surface (e.g.\ Cassinelli \& Olson 1979). Recently,
this idea was boosted by the direct measurement of magnetic fields on
some massive stars (e.g.\ Bouret \etal\ 2008a). Waldron \& Cassinelli
(2007) pointed out a ``high-ion near star problem'': the radii of
formation for the lines of ions with higher ionization potential ions
are closer to the surface than those of lower ions. Their proposed
explanation invokes magnetic fields.  Unfortunately, the
signal-to-noise ratio in the lines of He-like ions of Si, S, Ar, and
Ca, which provide the diagnostics for the hottest plasma, is rather
low. Better quality data are needed to pin-point the exact location of
the hot plasma in massive star winds in order to verify the claim of
Waldron \& Cassinelli.

However, surface magnetic fields may not be required to explain the
available measurements. The  simulations of instabilities in
stellar winds (Runacres \& Owocki 2002, 2005) show
that strong shocks may develop at the distances which agree well with
those inferred from the analysis of He-like ions.

Below, we briefly consider some solutions  proposed to
explain the observed emission line profiles:

\smallskip\noindent
{\it i)} Line optical depth alters the line shape of
X-ray emission profiles (Ignace \& Gayley 2002);

\smallskip\noindent
{\it ii)} 
Reduction of the wind absorption column density implying a lower
  $\dot{M}$ (e.g.\ Waldorn \& Cassinelli 2001; Cassinelli \etal\ 2001;
  Kramer, Cohen, \& Owocki 2003);

\smallskip\noindent {\it iii)}  Macroclumping resulting in smaller
effective opacity (e.g.\ Feldmeier, Oskinova, Hamann 2003; Oskinova, 
Feldmeier, Hamann 2006; Owocki \& Cohen 2006; Cassinelli \etal\ 2008).

\subsection{Optically thick X-ray emitting plasma}

Ignace \& Gayley (2002) calculated X-ray line profiles produced in an
hot plasma that is optically thick. They found that the optically thick
lines have nearly symmetrical shape. Leutenegger \etal\ (2007)
used this formalism to show that the resonance lines of He-like ions of
N and O in the X-ray spectrum of $\zeta$\,Pup are better described under
the assumption of resonant scattering.

\begin{figure}[h]
\centering
\includegraphics[width=7cm]{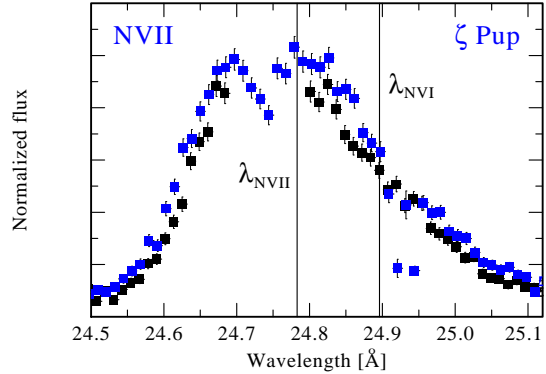}
\caption{The N\,{\sc vii} $\lambda$\,24.758 and N\,{\sc vi}
$\lambda$\,24.898 blend in the \xmm\ RGS1 (blue line) and RGS2 (red line) spectra 
of \zpup. The central wavelengths of corresponding lines are indicated.  
\label{fig:nvii}}
\end{figure}

{\changed{ Further evidence that some X-ray lines can be optically thick
comes from the characteristic shapes, e.g.\ the dip at the top of the
N\,{\sc vii}  line in $\zeta$\,Pup spectrum (Fig.\,\ref{fig:nvii}). 
Such line structures are typical for optically thick emission lines in
the optical spectra of WR stars, e.g.\ He\,{\sc
ii}\,$\lambda$\,5412\,\AA\ in WR3.}} As estimated by Ignace \& Gayley
(2002), the X-ray lines of leading ions in the most dense winds can
indeed be optically thick. However, the lines of less abundant ions
(such as S, Si, Ne, Fe) originating in the less dense winds cannot be
explained by resonant scattering. 

\subsection{Reducing the wind absorption column density}

The observed X-ray emission lines in O-star spectra are typically
symmetrical and similar across the X-ray spectrum. Fitting observed
lines with MacFarlane \etal\ line model, generally yields low values
of $\tau_0$ and its weak dependence on wavelength (e.g.\ Cassinelli
\etal\ 2001, Kahn \etal\ 2001, Miller \etal\ 2002, Kramer \etal\ 2003,
Pollock 2007). Recalling that $\tau_0\propto
\kappa_\lambda=\chi_\lambda\rho_{\rm w}$, these empiric results can be
explained by the weak wavelength dependence of opacity and by the
reduced wind density $\rho_{\rm w}$.

The wind opacity for the X-rays is mainly due to K-shell ionization of
metals. In O-stars $\kappa_\lambda$ chiefly depends on the chemical
composition, but little on other details of the wind models (Oskinova
\etal\ 2006). Hence, knowing the metal abundances is prerequisite to
calculate wind opacity.

\begin{figure}[h]
\begin{minipage}{9cm}
\centering
\includegraphics[width=9cm]{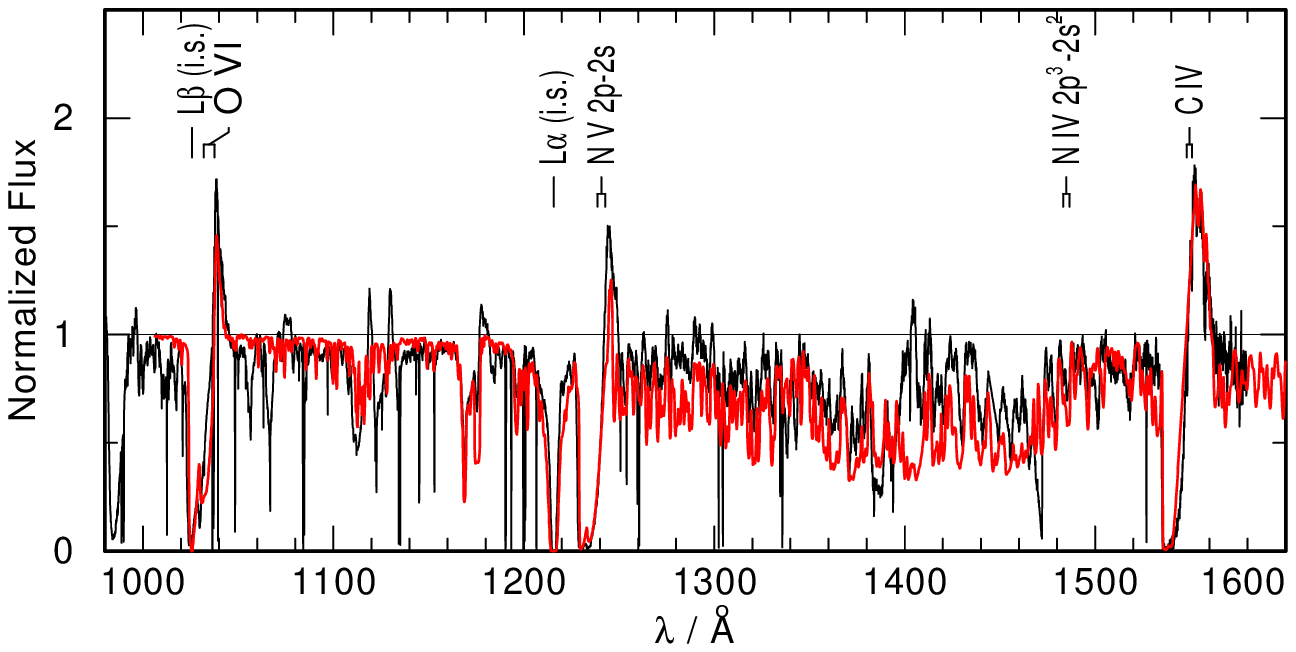}
\caption{The EUV spectrum of $\zeta$\,Pup, observed with FUSE (black
  line), compared to a PoWR model spectrum (red line). The
  resonance doublets of C\,{\sc iv}, N\,{\sc v} and O\,{\sc vi} are
  well reproduced, as well as the forest of iron-group lines. The 
  O\,{\sc vi} doublet can only be fitted with models accounting for
a  diffuse X-ray field. (Adopted from Oskinova \etal\ 2006). 
\label{fig:zpupuvsp}} 
\end{minipage}
\hfill
\begin{minipage}{7cm}
\centering
\includegraphics[width=7cm]{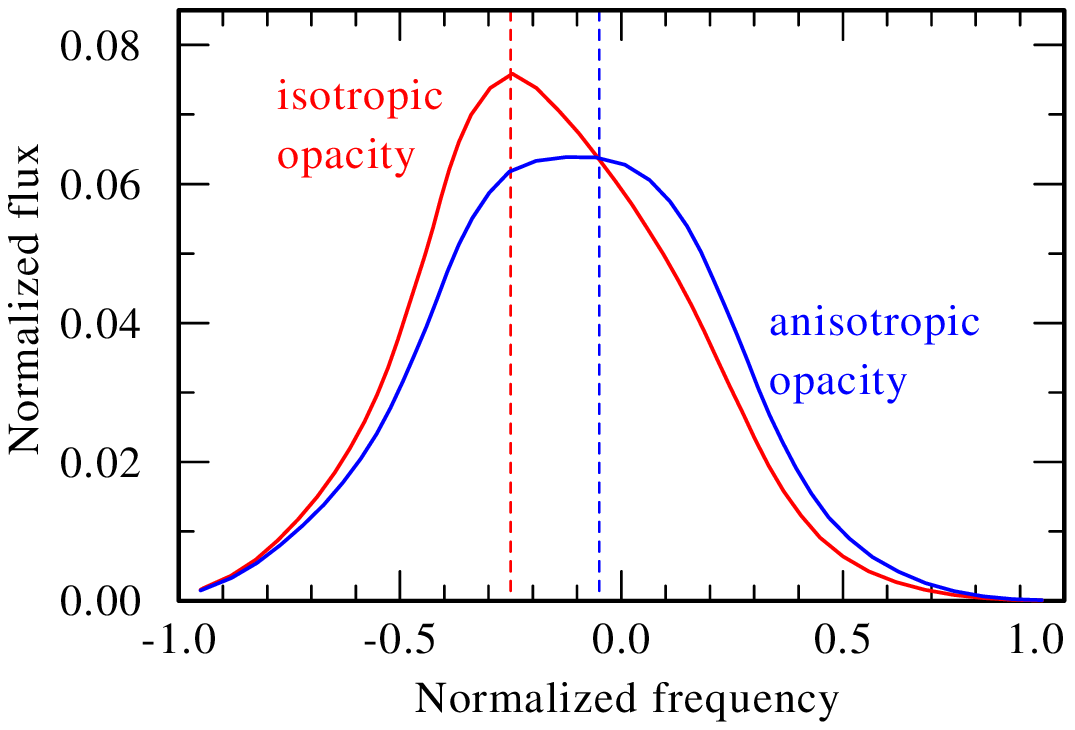}
\caption{
Two model lines calculated using the same wind parameters but assuming 
isotropic and anisotropic clumps. Isotropic opacity leads to a  
significantly more skewed line. \label{fig:flat}}
\end{minipage}
\end{figure}

The abundance in O-type stars are often non-solar (e.g.\ Lamers \etal
1999). Example of typical ON-type star with enhanced nitrogen
abundance is \zpup. Using non-LTE wind atmospheres, Pauldrach,
Hoffmann, \& Lennon (2001) found that while N is overabundant in this
star, C and O are depleted.  They show that the N/C ratio is
20$\times$solar and the N/O ratio is 10$\times$solar in
\zpup. Fig.\,\ref{fig:zpupuvsp} demonstrates our model fit to the UV
spectrum of $\zeta$\,Pup, assuming abundances as derived in Pauldrach
\etal. The lines of C, N, and O are well reproduced.

The abundances obtained from the analysis of X-ray spectra of
\zpup\ are in general agreement with those from Pauldrach
\etal\ (e.g.\ Fig.\,1, also Kahn \etal\ 2001).  Leutenegger
\etal\ (2007) estimate that nitrogen has twice the abundance of oxygen
in \zpup.  Krti\v{c}ka \& Kub\`at (2007) show that the use of new 3D
solar abundances (Asplund \etal\ 2005) with lower metalicity improves
the agreement between observation and wind theory.  Their calculations
of $\kappa_\lambda$ in $\xi$\,Per agree well with those in Oskinova
\etal\ (2006).

At softer X-ray energies the large wind optical depth for the X-rays
is largely determined by the CNO edges (see Fig.\,20 in Pauldrach
\etal).  Selectively reduced metal abundance would lead to a drop in
the jump heights in wind opacity due to the edges, leading
to less pronounced wavelength dependence of optical depth. Recently,
Cohen \etal\ (2010) assumed that in \zpup\ the abundance ratio of N/C
is 60$\times$solar and N/O is 25$\times$solar, which resulted in the
less steep jumps of wind optical depths at the wavelengths of the
K-shell edges. They claimed that this wind optical depth agrees with
the marginal wavelength dependence of $\tau_0$ (on 68\%\ confidence
limits) which they found by fitting the lines in the \cxo\ spectrum
of \zpup.  For consistency, we calculated the H$\alpha$ line adopting
the same wind parameters as used by Cohen \etal\ (2010) for their
X-ray line model. The resulting model H$\alpha$ line shows a large
discrepancy with the observed one. This points out that the parameters
used by Cohen \etal\ to model the X-ray lines may not be realistic.


The observed nearly symmetric shapes of the X-ray emission lines can
be explained if the wind density, $\rho_{\rm w}$, is small. The
reduction of the density can be expected from atmosphere models that
account for wind {\em clumping}.

The inhomogeneity of stellar winds has been established in the last
decade (see reviews in Hamann \etal\ 2008).  The mass-loss rates are
empirically inferred from modeling recombination and resonance
lines in the optical and UV. The recombination line strength depends on
$\rho_{\rm w}^2$, while the resonance line strength depends on
$\rho_{\rm w}$.  Assuming that the interclump medium is void and that
the density within the clumps is enhanced by a factor $D$, the
\mdot\ inferred from analysis of recombination lines has to be reduced
by factor $\sqrt{D^{-1}}$ compared to the values obtained under the
assumption of a smooth wind.

Assuming that nearly all mass is in clumps, the volume filling factor
is $f_{\rm V}\approx D^{-1}$. In order to fit recombination and
resonant lines simultaneously, the wind models require very small
filling factors (Fullerton, Massa, \& Prinja 2006).  Consequently,
these models require a reduction of empirically inferred \mdot\ up to a
factor of 100 compared to the unclumped models.

{\changed{Waldron \& Cassinelli (2010) proposed an alternative solution
to the problem of the mass-loss rate discordance. They pointed out that
the XUV radiation near the He\,{\sc ii} ionization edge originating in
wind shocks would destroy P\,{\sc v} ions. Consequently, the key 
diagnostic resonant line of P\,{\sc v} would be weakened and could be 
explained without the need to decrease  \mdot. The source of the XUV
radiation could be the bow-shocks around the wind clumps as proposed in 
Cassinelli \etal\ (2008). }}

The models with severely reduced mass-loss rates encounter substantial
problems when one tries to explain the observed X-ray line spectra:

\medskip\noindent {\it i)} The low values of \mdot\ do not explain why
the shapes of X-ray lines show no significant wavelength dependence.

\medskip\noindent {\it ii)} Eq.\,\ref{eq:t0} can be used to model the
X-ray lines only assuming that the clumps are optically thin.  This
strong assumption (often referred to as {\em microclumping}
approximation) is not always valid.

\section{Macroclumping}
\subsection{On the size of optically thick clumps}

There is no known reason why the microclumping approximation should
apply.  The optical depth of an isotropic clump (i.e. with the same
dimensions in 3D) is

\begin{equation}
\tau_{\rm clump}(\lambda)=\rho_{\rm w}\chi_{\lambda}D d_{\rm clump}\Rstar\,=\,
\frac{\mdot \chi_\lambda}{4\pi v_\infty \Rstar}\cdot\frac{d_{\rm clump}}{f_{\rm v}}
\cdot\frac{1}{(1-\frac{1}{r})^{\beta}r^2}\,\equiv\, \tau_\ast(\lambda)
\frac{d_{\rm clump}}{f_{\rm v}}
  \cdot\frac{1}{(1-\frac{1}{r})^{\beta}r^2},
\label{eq:tc}
\end{equation}
where  $d_{\rm clump}$\, is the geometrical size of the clump 
expressed in \Rstar. The strong wavelength dependence of $\chi_\lambda$ suggests 
that a clump may be optically thick at long wavelengths but thin at short ones.

It is convenient to express $\tau_\ast(\lambda)$ as
\begin{equation}
\tau_\ast(\lambda)\,\approx\,722\frac{\mdot_{-6}}{v_\infty {\cal R}_\ast}\cdot\chi_\lambda,
\end{equation}
where $\mdot_{-6}$  is the mass-loss rate in units
$10^{-6}$\,\myr, $v_\infty$ is the terminal velocity in [km\,s$^{-1}$],
and ${\cal R_\ast}=\Rstar/R_\odot$. If at some radius $r$, 
the clump size is larger than $d_{\rm clump}^{\tau=1}$, 
such clump is not optically thin for the X-ray radiation at $\lambda$. 
The size of a clump with optical depth $\tau=1$ at wavelength $\lambda$  
is
\begin{equation}
d_{\rm clump}^{\tau=1}=\frac{f_{\rm V}}{\tau_\ast(\lambda)}
\cdot\left(1-\frac{1}{r}\right)^{\beta}r^2. 
\label{eq:tc1}
\end{equation}
The microclumping approximation is valid only for significantly smaller
clumps.   

It is a common misconception to assume that {\em
  macroclumping}, which allows for any clump optical depth, implies
a geometrically large size of clumps.
Let us estimate the geometrical size of a clump which has optical
depth unity. We consider \zpup\ and use parameters from
Zsargo \etal\ (2008): $\mdot_{-6}=1.7$, $v_\infty=2300$, ${\cal R}=19$,
$\beta=0.9$. Thus, $\tau_\ast(\lambda)\,\approx\,0.03\chi_\lambda$.
Then a clump with optical depth unity has the size
\begin{equation}
d_{\rm clump}^{\tau=1}(\zeta\,{\rm Pup})=\frac{f_{\rm V}}{0.03}\cdot\frac{1}{\chi_\lambda}
\cdot\left(1-\frac{1}{r}\right)^{0.9}r^2. 
\label{eq:tc1}
\end{equation}
For a sample of O-stars, Bouret \etal\ (2008b) find $0.01 < f_\infty <
0.08$, where $f_\infty$ is the filling factor in wind regions where
$v_{\rm w}=\vinf$ (see Bouret \etal\ for details).  For a rough
estimate we adopt $f_{\rm V} = 0.03$. Then a clump of optical depth
unity has the size $d_{\rm clump}^{\tau=1}(\zeta\,{\rm
  Pup})=\chi_\lambda^{-1} \cdot(1-\frac{1}{r})^{0.9}r^2$.  Zsargo
\etal\ (2008) do not provide $\chi_\lambda$ values, and we are not
aware of any consistent calculation $\chi_\lambda$ for the low values
of \mdot.  Adopting $\dot{M}_{-6}=4.2$, Oskinova \etal\ (2006) derive
$\chi_\lambda\approx 60$ [cm$^2$\,g$^{-1}$] at 12\AA\ and 
$\chi_\lambda\approx 180$ [cm$^2$\,g$^{-1}$] at 19\AA. To roughly
account for the lower mass-loss rate, we reduce these values by a factor 
of two. The geometrical sizes of clumps which have optical depth unity at
12\AA\ and 19\AA\ in  \zpup\ wind are shown in Table\,1. 
Note that if \mdot\ is higher, clumps with even smaller geometrical sizes 
will be optically thick.

\begin{table}
\caption{Estimate of the geometrical size of a clump with optical
  depth unity at wavelength $\lambda$ in the wind of $\zeta$\,Pup. The wind
  parameters are from Zsargo \etal\ (2008) $\mdot_{-6}=1.7$,
  $v_\infty=2300$, $R=19$, $\beta=0.9$. 
Compared to Oskinova \etal\ (2006), $\chi_\lambda$ is scaled down by a
factor of two to account for the smaller \mdot\ adopted in Zsargo \etal\
(2008). 
\label{tab:cl}} 

\small
\begin{center}
\begin{tabular}{ | c  | c |  c | c | }
\hline  \hline
Wavelength $\lambda$ & $\chi_\lambda$ & \multicolumn{2}{c|}{$d_{\rm clump}(\tau_\lambda=1)\,[R_\ast]$ }
 \\ 
\hline
\AA\    &  [cm$^2$\,g$^{-1}$] &  At R = 2\Rstar\ in the wind & At R = 5\Rstar\ in the wind \\ 
\hline
12         & 30         & 0.07  & 0.7 \\
19         & 90         & 0.02  & 0.2 \\ 
\hline \hline
\end{tabular}
\end{center}
\end{table}

\noindent It is possible that the clumps in \zpup\ wind have sizes similar
to shown in Table\,1, and, thus, are optically thick at the
corresponding wavelengths.  Therefore the X-ray line profile fitting
based on the microclumping approximation can lead to the erroneous
results.

\subsection{Effective opacity}

The idea that clumping may reduce the wind opacity and lead to more
symmetric line profiles was briefly discussed in Waldron \& Cassinelli
(2001), Owocki \& Cohen (2001) and Kramer \etal\ (2003).  The effects of
wind clumping on the X-ray lines were investigated in detail in
Feldmeier \etal\ (2003), who solved the pure absorption case of
radiative transfer in clumped winds and found that the emission lines
are more symmetric than in the case of a smooth wind.  In Oskinova
\etal\ (2006) we employed a 2.5-D Monte-Carlo code (Oskinova \etal\
2004) to compute the  emission line profile for a finite number of
clumps and compared the  results to the observed lines.

The effective opacity,  $\kappa_{\rm eff}$,   in a clumped wind is the 
product of:
\begin{itemize}
\item average number of clumps per unit  volume, $n(r)$\,[cm$^{-3}$]
\item geometrical cross-section of a clump, $\sigma_{\rm clump}$\,[cm$^2$] 
\item probability that an X-ray photon with a wavelength $\lambda$ 
which encounters a wind clump gets absorbed,  
$\cal{P}$ = $1-{\rm e}^{-\tau_{\rm clump}(\lambda)}$ 
\end{itemize}
\noindent
Thus, 
\begin{equation}
\kappa_{\rm eff} = 
n(r)\cdot \sigma_{\rm clump} \cdot {\cal P} = 
\frac{\rho_{\rm w}\chi_{\lambda}}{\tau_{\rm clump}} \cdot (1-{\rm e}^{-\tau_{\rm clump}(\lambda)}), 
\label{eq:ke} 
\end{equation}
where we used Eq.\,(\ref{eq:tc}) and
expressed the filling factor as $f_{\rm V}=D^{-1}=n(r)V_{\rm clump}$
with $V_{\rm clump}=\sigma_{\rm clump} d_{\rm clump}R_\ast$. In the case
of optically thin clumps ($\tau_{\rm clump}\ll 1$) the effective
opacity is $\kappa_{\rm eff} = n(r)\cdot \sigma_{\rm
  clump}\cdot\tau_{\rm clump}=\rho_{\rm w}\chi_{\lambda}\equiv
\kappa_\lambda$, recovering the microclumping approximation. In the case of
optically thick clumps ($\tau_{\rm clump}\gg 1$), the absorption probability is
${\cal P}=1$, yielding $\kappa_{\rm eff}=n(r)\sigma_{\rm clump}$. In this limit
of porous wind,
the opacity does not depend of the wavelength, but is ``gray''.

In general, the effective opacity is wavelength dependent.  As can be
seen from Eqs.\,(\ref{eq:tc}) and (\ref{eq:ke}), the dependence on
wavelength enters via the clump optical depth:  $\kappa_{\rm eff}\propto
1-{\rm e}^{-\tau_{\rm clump}(\lambda)}$.  This reduced  dependence of effective
opacity on wavelength agrees well with observations.

Evaluating the wind optical depth as an integral over effective opacity
along the line-of-sight $z$: 

\begin{equation} 
\tau_{\rm w}=\int_{z_\nu}^\infty n(r)\sigma_{\rm clump}
(1-{\rm e}^{-\tau^{\rm clump}_\nu}) {\rm d}z,  
\label{eq:tol} 
\end{equation} 

\noindent where $z$ is the coordinate along the line of sight. Motivated
by the results of the hydrodynamic simulations which predict radially
compressed wind structures, Feldmeier \etal\ (2003) studied the case of
radially compressed clumps, with $\sigma_{\rm clump}\propto |dr/dz|$. 
In this case, the integral in Eq.\,(\ref{eq:tol}) transforms into an
integral over $r$.  As a result, the X-ray emission line profiles will
be nearly symmetric (see Fig.\,4), while in the case of isotropic clumps
the line profiles are more skewed (see also Herv{\'e} \& Rauw 2011,
these proc.). Hence, lower values of $\tau_0$ are obtained when the
X-ray lines are fitted using the specific assumption of spherical
clumps compared to the case of angular dependent opacity.

\subsection{The impact of clumping on empirical mass-loss rate estimates}

Owocki, Gayley, \& Shaviv (2004) studied the effects of porosity on
the atmospheres of LBV stars. Massa \etal\ (2003) and Fullerton
\etal\ (2006) discussed how porosity can affect the formation of
P\,Cygni lines.  Prinja \& Massa (2010) found the spectroscopic
signatures of wind clumping, and show that macroclumping must be taken
into account to model the UV resonance lines. The effect of
macroclumping on {\em line opacity} was studied for the first time in
Oskinova, Hamann, \& Feldmeier (2007). The macroclumping was
incorporated in the state-of-the-art non-LTE atmosphere model PoWR 
(e.g.\ Gr\"afener, Koesterke \& Hamann 2002).  We have shown that
accounting for clumps leads to empirical mass-loss rate estimates
which are by a factor of a few higher than those obtained under the
assumption of microclumping, but are still reduced compared to a
smooth wind.

\begin{figure}[h]
\begin{minipage}{7cm}
\centering
\includegraphics[width=6cm,
angle=-90]{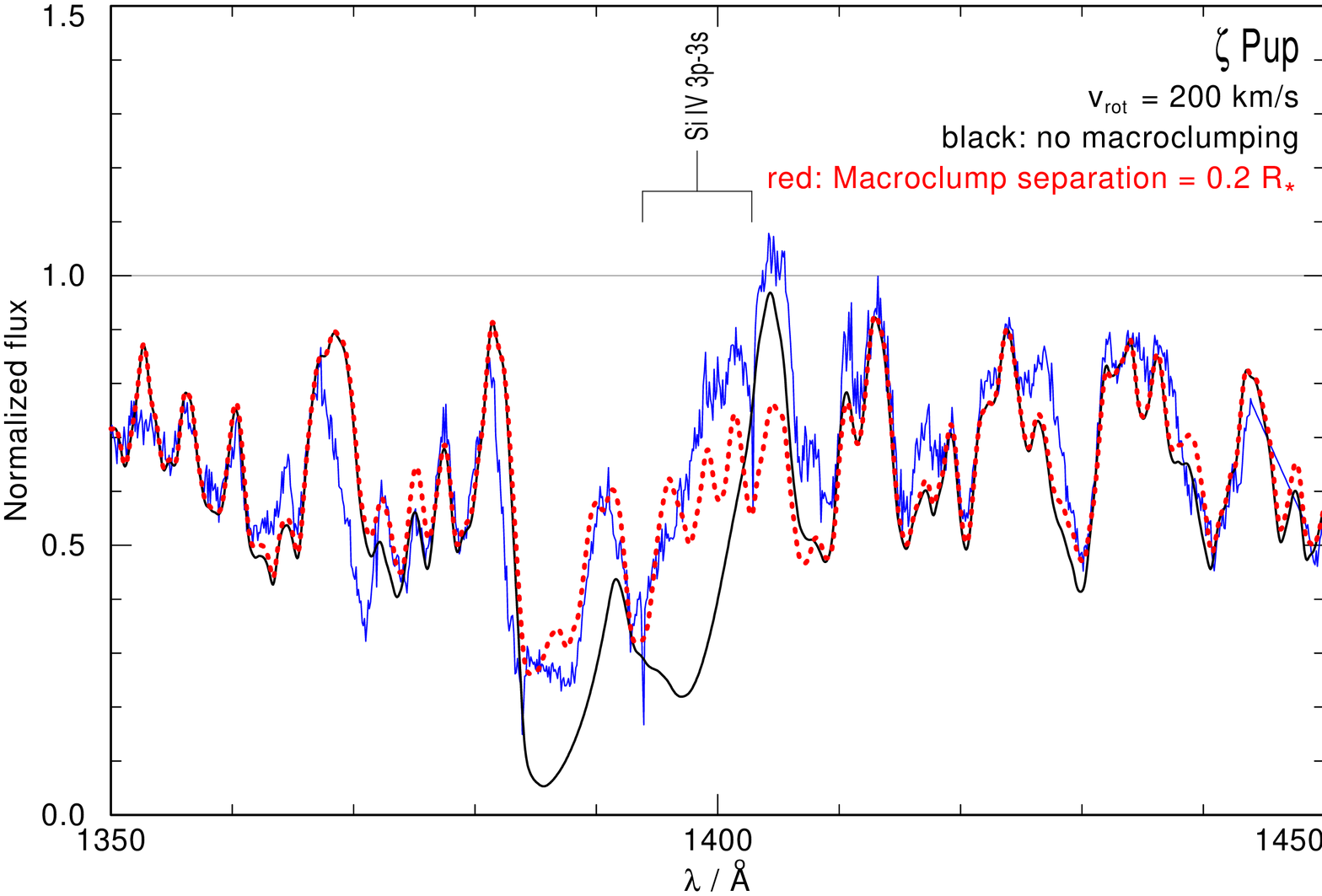}
\caption{
Effect of macroclumping on the Si\,{\sc iv} doublet. The IUE
spectrum of $\zeta$\,Pup is shown in blue.  The
usual microclumping modeling yields P\,Cygni features that are too
strong (black, continuous line). With our macroclumping formalism, the
line features are reduced to the observed strength (red, dotted
curve). 
\label{fig:si}}
\end{minipage}
\hfill
\begin{minipage}{7cm}
\centering
\includegraphics[width=7cm]{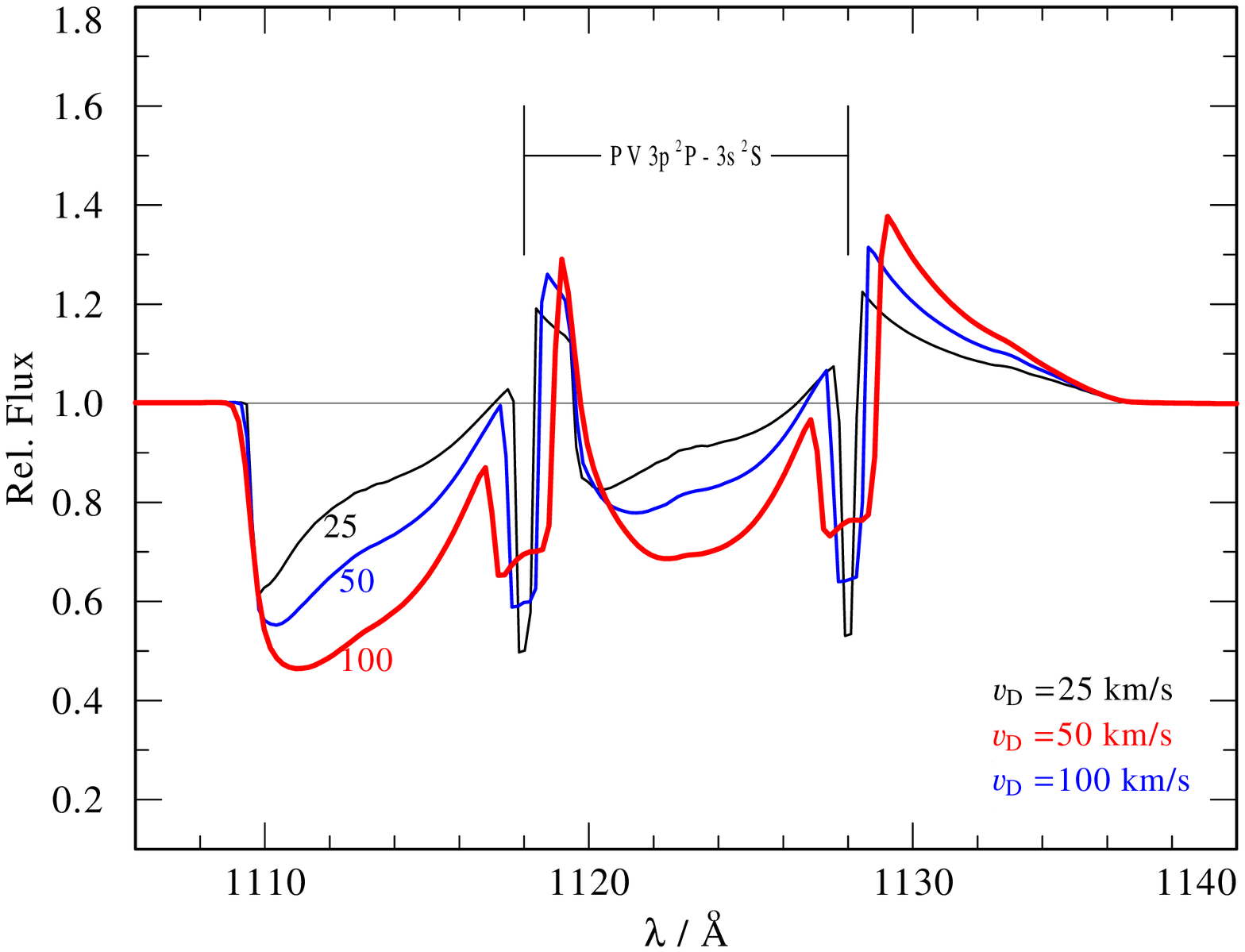}
\caption{Synthetic P\,{\sc v} resonance doublet at
  $\lambda$\,1118/1128\,\AA\ for different values of the
  microturbulence velocity $v_{\rm D}$ (labels). All other model
  parameters are kept fixed. When the velocity dispersion across the
  clumps is decreased, the macroclumping effect becomes more
  pronounced and leads to a weaker line profile. \label{fig:vd}}
\end{minipage}
\end{figure}

In the case of line opacity, an additional parameter, $v_{\rm D}$ is
required, which describes the velocity field {\em within} the clump.
The detailed UV line fits for O stars require a "microturbulence
velocity'' of 50 to 100 km\,s$^{-1}$, which, perhaps, can be
attributed to the velocity dispersion within clumps.
Figures\,\ref{fig:si} and \ref{fig:vd} illustrate the effect of
macroclumping on resonance lines. With a lower velocity dispersion
within a clump, the Doppler broadening becomes smaller, and the line
absorption profile is narrower but peaks higher. Thus, the clump
optical depth becomes larger in the line core, but smaller in the line
wings. In the statistical average, this leads to a reduction of the
effective opacity and a weakening of the line (see also Sundqvist
\etal\ 2011, these proc.).

At present macroclumping is included in the non-LTE model atmosphere
only as a first approximation. Even though the results are very
encouraging because consistent mass-loss rates can be obtained
simultaneously from the analysis of UV, optical, and X-ray spectra. As
an example, our analyses of H$\alpha$, P\,{\sc v}, and Si\,{\sc iv}
lines  and the X-ray emission lines all agree with a value
for the mass-loss rate of $\mdot \approx 2.5 \times 10^{-6}$\,\myr\ in
the wind of $\zeta$\,Pup (Oskinova \etal, in prep.).

\section{The large scale structures in stellar winds and the X-rays}

\subsection{X-rays and  Co-rotating Interaction Regions}

In the previous section, we discussed the small-scale, stochastic wind
inhomogeneities. Beside those, there is strong evidence for the presence
of large-scale structures in stellar winds. {\changed{Spectral lines
formed in stellar winds are variable}}, e.g.\ discrete absorption
components (DACs) are observed in the UV resonance lines of nearly all O
stars (Prinja \& Howarth, 1986).  Cranmer \& Owocki (1996) explained
DACs as originating from co-rotating interaction regions (CIR), where
high-density, low-speed streams collide with low-density, high-speed
streams. The observed slow drift of DACs can be understood by
considering the motion of the patterns in which the DAC features are
formed (Hamann \etal\ 2001). The hydrodynamics of stellar winds which
can explain the DACs and the faster modulations are considered by Lobel
\etal\ (2011, these proc.).

This complex wind geometry should affect the production and the
propagation of X-rays in stellar winds: the wind can be shocked at the
CIR surfaces (Mullan 1984). X-rays may suffer additional absorption
in density enhanced CIRs.

The DACs recurrence time is on time scale of days. Oskinova, Clarke,
\& Pollock (2001) detected periodic X-ray variability with an
amplitude of $\sim$20\%\ in the ASCA passband (0.5-10 keV) of the O9Ve
star $\zeta$\,Oph. The detected period of $0\hbox{$.\!\!^{\rm d}$}77$
possibly indicates a connection with the recurrence time
($0\hbox{$.\!\!^{\rm d}$}875 \pm 0\hbox{$.\!\!^{\rm d}$}167 $) of the
DACs in the UV spectra of this star. In contrast, the analysis of
$1\hbox{$.\!\!^{\rm d}$}175$ continuous ASCA observations of
$\zeta$\,Pup failed to confirm the previously reported variability at
the 6\%\ level with a period of 16.667\,h found in earlier Rosat data.
The new, more sensitive analyses of X-ray observations of O-stars
which will extend over the stellar rotational periods should shed more
light on the connections between DACs, rotation, and X-ray emission.

\subsection{X-ray emission from Oe-type stars}

$\zeta$\,Oph belongs to the rare class of Oe stars (Negueruela, Steele, 
\& Bernabeu 2004).  Oe-type stars display Balmer lines similar to
those in the classical Be stars. The latter are fast rotating stars with
decretion disks.  Negueruela \etal\  point out that the Oe-phenomenon is
restricted to the latest subtypes among the O stars, indicating that the
building of disks is more problematic for the higher-mass stars.  

Li \etal\ (2008) studied the X-ray emission from Oe/Be stars to test
whether the disks of these stars could form by magnetic channeling of
wind toward the equator {\changed{(Cassinelli \etal\ 2002)}}. In their
model, X-rays can be produced by  material that enters the shocks above
and below the disk region. The model by Li \etal\ predicts an existence
of a relation between  \Lx/\Lbol\ and the magnetic field strength in
Oe/Be stars.

High-resolution X-ray spectra are only available for two Oe stars:
$\zeta$\,Oph (O9Ve) and HD\,155806 (O7.5Ve). The X-ray properties of
these stars are quite different.  $\zeta$\,Oph shows periodic
modulations of the X-ray flux, has narrow X-ray emission lines. The bulk
of its plasma is at a high temperature of 8\,MK (Zhekov \& Palla 2004).
Naz\'e \etal\ (2010a) do not detect a modulation of the X-ray flux in
HD\,155806 and report that its X-ray emission lines are broad. The bulk
of its hot plasma has only 2\,MK. The $\log{\Lx/\Lbol}=-6.75$ for
HD\,155806, while $\log{\Lx/\Lbol}=-7.4$ for $\zeta$\,Oph (Oskinova
\etal\ 2006, Naz\'e \etal\ 2010a). Clearly, larger observational samples
are required to understand the link between the X-ray emission of
Oe-type stars and their hypothetical circumstellar disks.

\subsection{X-ray emission from O-stars with magnetic fields}

Large-scale flow structures in stellar winds can result when a
large-scale magnetic field confines the outflow of matter. Babel \&
Montmerle (1997) studied the case of a rotating star with a sufficiently
strong dipole magnetic field. Collision between the wind components from
the two hemispheres in the closed magnetosphere leads to a strong shock
and X-ray emission. Based on this magnetically confined wind shock model
(MCWS), the presence of a magnetic field on the O-type star
$\theta^1$\,Ori\,C had been postulated. Direct confirmation of the
magnetic field in this star by Donati \etal\ (2002) proved that X-rays
have large diagnostic potential in selecting massive stars with surface
magnetic fields.

The MHD simulations in the framework of the MCWS model were
performed by ud-Doula \& Owocki (2002) and Gagn{\'e} \etal\ (2005).
Using as input parameters the characteristic values of the wind and
the magnetic field strength of $\theta^1$\,Ori\,C, these simulations
predict the plasma temperature, emission measure, and periodic
X-ray flux modulation which compare well with
observations.

This modeling success established the MCWS model as a general scenario
for the X-ray emission from magnetic early type stars. The MCWS model
makes predictions that can be directly compared with observations:
{\it i)} the hottest plasma should be located at a few stellar radii
from the stellar surface at the locus where the wind streams collide;
{\it ii)} the X-ray emission lines should be rather narrow, because
the hot plasma is nearly stationary; {\it iii)} magnetic stars should
be more X-ray luminous than their non-magnetic counterparts of similar
spectral type; {\it iv)} the X-ray spectrum of magnetic stars should
be harder than that of non-magnetic stars, with the bulk of the hot
plasma at temperatures $\sim$20\,MK; {\it v)} the X-ray emission
should be modulated periodically as a consequence of the occultation
of the hot plasma by a cool torus of matter, or by the opaque stellar
core. X-ray variability may be expected when the torus breaks up.

The X-ray observation of magnetic O-type stars led to perplexing results
that are not always in agreement with the model predictions.
$\zeta$\,Ori\,A has a weak surface magnetic field, that may be
responsible for the presence of hot plasma close to stellar surface
(Waldron \& Cassinelli 2007). But, in general, its X-ray properties are
typical for an O-type star (Raassen \etal\ 2008). A strong magnetic
field ( $\sim 1$\,kG) is detected on HD\,108 (O7I) (Martins \etal\
2010). However, the emission measure (EM) of the softer spectral
component, with a temperature of $\approx 2$\,MK, is more than one order
of magnitude higher than the EM of the harder component $T_{\rm
max}\approx 15$\,MK, contrary to the expectation of the MCWS model
(Naz{\'e} \etal\ 2004).  HD\,191612 also has a $\sim 1$\,kG strong
magnetic field (Donati \etal\ 2006a). Recently, Naz{\'e} \etal\ (2010b)
demonstrated that the large EM at $\approx$2\,MK and the broad X-ray
emission lines in the X-ray spectrum of this star do not compare well
with the predictions of the MCWS model. The early-type B-star
$\tau$\,Sco has a complex magnetic field topology (Donati \etal\ 2006b)
and a hard X-ray spectrum (Wojdowski \& Schulz 2002, Mewe \etal\ 2003). 
A substantial modulation of the X-ray flux with stellar rotation period
was expected, but indications for only marginal variability were found
by Ignace \etal\ (2010).

Overall, considering the analysis of X-ray observations of magnetic O
stars, it appears that only one star, $\theta^1$\,Ori\,C, displays the
properties that are fully compatible with the MCWS model.

\subsection{X-ray emission from WR stars} The X-ray emission from WR-type
stars remains enigmatic -- some WN-type stars are X-ray sources
(Ignace \etal\ 2003, Skinner \etal\ 2010), while others remain
undetected despite low upper limits on $L_{\rm X}$ (Gosset \etal\
2005).  Oskinova \etal\ (2003) showed that WC-type stars are not X-ray
sources, a result which they attribute to the very large wind opacity.
WO-type star winds are even more metal enriched. {\changed{However, a
reduction in the mass-loss (a poorly constrained parameter) by a factor
of only two and/or a higher effective stellar temperature result in a
higher degree of wind ionization. In this case a fraction of X-rays
could escape. Wind  anisotropy can further mitigate wind attenuation.}}

Magnetic fields and large-scale distortions of stellar winds are invoked
as possible explanations for the recently detected X-ray emission from
the WO-type star WR\,142 (Oskinova \etal\ 2009). WR\,142 is a massive
star in a very advanced evolutionary stage shortly before its explosion
as a supernova or $\gamma$-ray burst. From qualitative considerations we
conclude that the observed X-ray radiation is too hard to allow
wind-shock origin of X-ray emission. The proposed explanation of its
X-ray emission suggests surface magnetic field. Possibly related, WR 142
seems to rotate extremely fast, as indicated by the unusually round
profiles of its optical emission lines. Our X-ray detection implies that
the wind of WR\,142 must be relatively transparent to X-rays, which
could be due to strong wind ionization, wind clumping, or non-spherical
geometry from rapid rotation.

\section{Open questions}
We find that incorporating macroclumping in the wind models allows to
explain the shapes of X-ray emission lines in O-star spectra.
However, many questions about X-rays from single massive stars
remain. We list a subjective selection of these questions, which
we think are the most promising ones to answer with new advances in
theory and observations:

\smallskip\noindent
-- Is there a correlation between $T_{\rm X}$ and $T_{\rm eff}$ as found by 
Walborn, Nichols, \& Waldron (2009)?

\smallskip\noindent
-- Is there a near star high ion problem? Is there a dependence of $\tau_0$ on 
the radius of line formation? 

\smallskip\noindent
-- What is the origin of the $\Lx \propto \Lbol$ correlation and how to
  explain deviations from it?

\smallskip\noindent
-- Does the MCWS model explains the different X-ray properties of magnetic OB-stars?

\smallskip\noindent
-- Why  X-rays from Oe/Be stars are not meeting the model expectations?

\smallskip\noindent
-- How X-rays are produced in WR-stars? 

\section*{Acknowledgements}
Authors are grateful to J.\,P.~Cassinelli for the insightful
discussion.

%
%
\footnotesize
\beginrefer
\refer Asplund, M., Grevesse, N., \& Sauval, A.J. 2005, ASP Conf. Ser., 336, 25

\refer Babel, J., \& Montmerle, T. 1997, A\&A, 323, 121


\refer Baum, E., Hamann, W.-R., Koesterke, L., Wessolowski, U., 1992, 
A\&A, 266, 402

\refer Bouret, J.-C., Donati, J.-F., Martins, F., \etal, 2008a, MNRAS,
389, 75

\refer Bouret, J.-C, Lanz, T., Hillier, D., \etal, 2008b, 
Clumping in Hot Star Winds, Universitatsverlag, Potsdam, p.31
	

\refer Cassinelli, J.P. \& Olson, G.L., 1979, ApJ 229, 304

\refer Cassinelli, J.P. \&  Swank, J.H., 1983, 271, 681 

\refer 	Cassinelli, J.P.,  Miller, N.A., Waldron, W.L., MacFarlane, J.J.,
 Cohen, D.H. \etal, 2001, ApJ, 554, 55

\refer Cassinelli, J.P., Brown, J.C., Maheswaran, M., Miller, N.A.,
 Telfer, D.C., 2002, ApJ, 578, 951

\refer Cassinelli, J.P., Ignace, R., Waldron, W.L., Cho, J., 
Murphy, N.A., Lazarian, A. 2008, ApJ, 683, 1052
	
\refer Cohen, D., Leutenegger, M. A., Wollman, E.E., \etal\ 2010, 
MNRAS, 405, 2391

\refer 	Cranmer, S.R. \& Owocki, S.P., 1996, ApJ, 462, 469

\refer Donati, J.-F., Babel, J., Harries, T. J., \etal, 2002, MNRAS, 333, 55

\refer Donati, J.-F., Howarth, I. D., Bouret, J.-C., Petit, P., 
Catala, C., Landstreet, J., \etal, 2006a, MNRAS, 365, L6

\refer Donati, J.-F., Howarth, I.D.; Jardine, M.M., 
\etal, 2006b, MNRAS, 370, 629

\refer Gagn{\'e}, M., Oksala, M.E., Cohen, D.H. \etal\ 2005, ApJ, 628, 986

\refer Gosset, E., Naz\'e, Y., Claeskens, J.-F., Rauw, G., Vreux, J.-M.,
 Sana, H. 2005, A\&A, 429, 685

\refer 	G\"udel, M. \&  Naz\'e, Y., 2009, A\&ARv, 17, 309

\refer  Gr\"afener, G., Koesterke, L.,\& Hamann, W.-R, 2002, A\&A, 387, 244

\refer Feldmeier, A., Puls, J., Pauldrach, A. W. A., 1997, A\&A, 320, 899
	
\refer 	Feldmeier, A., Oskinova, L., Hamann, W.-R., 2003, A\&A, 403, 217

\refer Fullerton, A. W., Massa, D. L., \& Prinja, R. K. 2006, ApJ, 637, 1025
	
\refer Hamann, W.-R., Brown, J.C., Feldmeier, A., Oskinova, L.M. 2001, A\&A, 
378, 946

\refer	Hamann, W.-R. \etal, 2008, Clumping in Hot Star Winds. 
Universitatsverlag, Potsdam

\refer Herv{\'e}, A. \& Rauw, G., 2011, LIAC2010

\refer	Hillier, D. J., Kudritzki, R.P., Pauldrach, A.W., \etal,
 1993, A\&A, 276, 117
 
\refer Ignace, R., 2001, ApJ, 549, 119

\refer 	Ignace, R., Oskinova, L. M., \& Brown, J. C.,  2003, A\&A, 408, 353

\refer 	Ignace, R. \& Gayley, K. G., 2002, ApJ, 568, 954 

\refer 	Ignace, R., Oskinova, L. M., Jardine, M. \etal, 2010, ApJ, 721, 1412
	
\refer 	Kahn, S.M., Leutenegger, M. A., Cottam, J., \etal\ 2001, A\&A, 365, 312

\refer	Kramer, R.H., Cohen, D.H., Owocki, S.P., 2003, ApJ 592, 532

\refer	Krti\v{c}ka, J. \& Kub{\'a}t, J., 2007, A\&A, 464, 17 


\refer	Lamers, H.J.G.L.M., Haser, S., de Koter, A., Leitherer, C., 
1999, ApJ, 516, 872

\refer 	Leutenegger, M.A., Paerels, F.B.S, Kahn, S.M., Cohen, D.H.,  
2006, ApJ, 650, 1096

\refer Leutenegger, M.A., Owocki, S.P., Kahn, S.M., \& Paerels, F.B.S. 2007, 
ApJ, 659, 642

\refer Li, Q., Cassinelli, J.P., Brown, J.C., Waldron, W.L., Miller, N.A.,  
2008, ApJ, 672, 1174

\refer Lobel, A. \etal, 2011, LIAC2010



\refer	MacFarlane, J.J., Cassinelli, J.P., Welsh, B.Y., Vedder, P.W., 
Vallerga, J. V., Waldron, W. L. 1991, ApJ 380, 564

\refer Martins, F., Donati, J.-F., Marcolino, W.L.F., \etal, 2010, 
MNRAS, 407, 1423

\refer Massa, D. Fullerton, A.W., Sonneborn, G., Hutchings, J. B.,
2003, ApJ, 586, 996

\refer Mewe, R., Raassen, A.J.J., Cassinelli, J.P., van der Hucht, K. A.,
 Miller, N. A., G\"udel, M., 2003, A\&A, 398, 203
	
\refer	Miller, N.A., Cassinelli, J.P., Waldron, W.L., MacFarlane, J.J., 
Cohen, D.H., 2002, ApJ, 577, 951

\refer Mullan, D.J., 1984, ApJ 283, 303

\refer Naz{\'e}, Y.; Rauw, G.; Vreux, J.-M.; De Becker, M. 2004, A\&A, 417, 667

\refer Naz{\'e}, Y., 2011,  LIAC2010

\refer Naz\'e, Y.; Rauw, G.; Ud-Doula, A., 2010a, A\&A, 510, 59

\refer Naz{\'e}, Y., ud-Doula, A., Spano, M., Rauw, G., De Becker, M., 
Walborn, N.R., 2010b, A\&A, 520, 59


\refer	Negueruela, I., Steele, I.A., Bernabeu, G., 2004, AN, 325, 749

\refer	Oskinova, L.M., Clarke, D., Pollock, A.M.T., 2001, A\&A, 378, L21

\refer Oskinova, L.M., Ignace, R., Hamann, W.-R., Pollock, A.M.T., 
Brown, J.C., 2003, A\&A, 402, 755

\refer	Oskinova, L.M., Feldmeier, A., Hamann, W.-R.,  2004, A\&A, 422, 675

\refer  Oskinova, L.M.; Feldmeier, A.; Hamann, W.-R., 2006,  MNRAS, 372, 313

\refer	Oskinova, L.M., Hamann, W.-R., Feldmeier, A., 2007,  A\&A, 476, 1331

\refer Oskinova, L.M, Hamann, W.-R., Feldmeier, A., Ignace, R., 
Chu, Y.-H., 2009, ApJL, 693, 44

\refer Owocki, S. P., Castor, J. I., Rybicki, G. B., 1988, ApJ, 335, 914

\refer Owocki, S.P. \& Cohen, D. H., 2001, ApJ, 559, 1108

\refer Owocki, S.P., Gayley, K.G., Shaviv, N.J., 2004, ApJ, 616, 525 

\refer Owocki, S.P. \& Cohen, D. H., 2006, ApJ, 648, 565 

\refer Pauldrach, A.W.A,  Hoffmann, T.L, Lennon, M., 2001, A\&A 375, 161

\refer 	Pollock, A. M. T., 2007, A\&A, 463, 1111


\refer  Prinja, R.K. \& Howarth, I.D., 1986, ApJS, 61, 357

\refer	Prinja, R.K \& Massa, D., 2010, A\&A, 521, L55

\refer	Raassen, A.J.J., van der Hucht, K.A., Miller, N.A., Cassinelli, J.P., 2008, A\&A, 478, 513

\refer	Runacres, M. C. \& Owocki, S. P., 2002, A\&A, 381, 1015

\refer	Runacres, M. C. \& Owocki, S. P., 2005, A\&A, 429, 323

\refer Skinner, S.L., Zhekov, S.A., G\"udel, M., Schmutz, W., Sokal, 
K.R., 2010, AJ, 139, 825

\refer Sundqvist, J., \etal, 2011, LIAC2010



\refer	Walborn, N.R., Nichols, J. S., Waldron, W.L., 2009, ApJ, 703, 633

\refer Waldron, W. L., 1984, ApJ, 282, 256
	
\refer	Waldron, W. L. \& Cassinelli, J. P., 2001, ApJ, 548, 45 

\refer	Waldron, W. L. \& Cassinelli, J. P., 2007, ApJ, 668, 456 

\refer	Waldron, W. L. \& Cassinelli, J. P., 2010, ApJ, 711, 30
 	
\refer	Wojdowski, P.S \& Schulz, N.S, 2002, ApJ, 627, 953

\refer ud-Doula, A., \& Owocki, S. P. 2002, ApJ, 576, 413
	
\refer Zhekov, S.A. \& Palla, F. 2004, 	MNRAS, 382, 1124
	
\refer	Zsarg{\'o}, J., Hillier, D.J., Bouret, J.-C., Lanz, T., 
Leutenegger, M.A., Cohen, D.H., 2008, ApJ, 685, 149

\endrefer           

\end{document}